\documentclass[manuscript,screen]{acmart}
\usepackage{xcolor}
\usepackage{color,soul}
\usepackage{cite}
\usepackage{amsmath}
\usepackage{pdfpages}

\AtBeginDocument{%
  \providecommand\BibTeX{{%
    \normalfont B\kern-0.5em{\scshape i\kern-0.25em b}\kern-0.8em\TeX}}}

\setcopyright{none}
\renewcommand\footnotetextcopyrightpermission[1]{} 
\acmConference[Under Review]{Make sure to enter the correct
 conference title from your rights confirmation email}{2022}
 
\begin{document}

\title{Explainability in Machine Learning: a Pedagogical Perspective}



\author{Andreas Bueff}
\affiliation{%
 \institution{University of Edinburgh}
  \country{United Kingdom}
}

\author{Ioannis Papantonis}
\affiliation{%
 \institution{University of Edinburgh}
  \country{United Kingdom}
 }


\author{Auste Simkute}
\affiliation{%
 \institution{University of Edinburgh}
  \country{United Kingdom}
  }

\author{Vaishak Belle}
\affiliation{%
 \institution{University of Edinburgh}
  \country{United Kingdom}
  }



\begin{abstract}
Given the importance of integrating of explainability into machine learning, at present, there are a lack of pedagogical resources exploring this. Specifically, we have found a need for resources in explaining how one can teach the advantages of explainability in machine learning. Often pedagogical approaches in the field of machine learning focus on getting students prepared to apply various models in the real world setting, but much less attention is given to teaching students the various techniques one could employ to explain a model's decision-making process. Furthermore, explainability can benefit from a narrative structure that aids one in understanding which techniques are governed by which questions about the data.

We provide a pedagogical perspective on how to structure the learning process to better impart knowledge to students and researchers in machine learning, when and how to implement various explainability techniques as well as how to interpret the results. We discuss a system of teaching explainability in machine learning, by exploring the advantages and disadvantages of various opaque and transparent machine learning models, as well as when to utilize specific explainability techniques and the various frameworks used to structure the tools for explainability. Among discussing concrete assignments, we will also discuss ways to structure potential assignments to best help students learn to use explainability as a tool alongside any given machine learning application. 

Data science professionals completing the course will have a birds-eye view of a rapidly developing area and will be confident to deploy machine learning more widely. A preliminary analysis on the effectiveness of a recently delivered course following the structure presented here is included as evidence supporting our pedagogical approach.
\end{abstract}



\keywords{Machine learning, data analysis, explainable artificial intelligence, interpretable data models }

\maketitle

\section{Introduction}


At present, machine learning models have demonstrated success on numerous tasks and challenges, both supervised and unsupervised. As technological advancements have increased computational hardware, modern research has resulted in numerous high performing machine learning methods. In turn, many opaque machine learning implementations are implemented in the modern world. Machine learning (ML) has many applications, which operates usually as a black box, where output of the model in no way indicates how it came to a decision.  

In order to trust black-box models, certain assurances need to be in place for further global implementation. Areas such as fairness, reliability, safety, explanatory justifiability, privacy, usability, and plenty more all are requisites for future models. Such criteria pose a challenge to continued research in areas of machine learning. Explainability and interpretability has particular relevance.

 In the context of machine learning, explainability refers to the collection of features of structured domains that have contributed for a given example to produce a decision. 
The success of various black-box models, along with the various fields in which they are implemented, has also come with notable challenges: how can we understand the decisions suggested by these models in order to trust them? As black-box models are inherently opaque, understanding their inner decision making process is non-trivial, and so human trust in such systems comes with inherent risk. The application of explainability techniques can facilitate a better understanding of various aspects of a model and its reasoning, leading to a better measure of trust on the part of various stakeholders.

We provide a pedagogical perspective on how to structure the learning process to better impart knowledge to students and researchers in machine learning, when and how to implement various explainability techniques as well as how to interpret the results. We discuss a system of teaching explainability in machine learning, by exploring the advantages and disadvantages of various opaque and transparent machine learning models, as well as when to utilize specific explainability techniques and the various frameworks used to structure the tools for explainability. Among discussing  assignments, we will also discuss ways to structure potential assignments to best help students learn to use explainability as a tool alongside any given machine learning application. 


These notions have been concretely structured into a recently deployed course at the authors' university, to both students and industry professionals. Discussions below reflect our current practice. Thus, it is important to note that the following material largely  elaborates on our current deployment strategy but is not to be seen as a dogmatic regime that must be followed. Rather, our pedagogical approach can be interpreted as a template for teaching explainability in machine learning.

The course focuses on understanding and applying various explainable machine learning techniques and judging their shortcomings. The course emphasises methodological thinking and understanding principles for the application of post-hoc explainability. The course is based on comprehensive recent surveys \citep{belle2020principles,ARRIETAxai, molnar2019}. Regarding terminology in this report, we abbreviate the referencing of explainable machine learning as eXplainable Artificial Intelligence (XAI) , which refers to the broader area of research concerned with producing explainable models without sacrificing performance and building trust between humans and black-box models through better interfacing \citep{GunningExplainableXAI}. While the techniques covered in this report represent only a limited perspective of XAI, we maintain a scoped reference to XAI through out this discussion for general simplicity with the understanding that the term can be applied more broadly, especially because machine learning and data science is but a subset of artificial intelligence \citep{ChakrabortiSZK17,kulkarniExplicablePlanning}. Furthermore, we note that explainability and interpretability are terms frequently used interchangeably in XAI literature, however there are specific differences in the concepts. Interpretability is defined by a passive characteristic of a model to make sense for a human observer, and explainability is defined as an active characteristic of a model where the model takes deliberate actions to explain its internal reasoning for the sake of human clarity \citep{ChakrabortiSZK17, RibeiroSG16}. For the sake of simplicity, we drop the distinction between the two terms for ease of readability of this report, and to a large degree the methods covered in this course fall under interpretability.

The course explores the principles and practice of enabling explainability in machine learning models. The course builds a narrative around a putative data scientist, and discusses how she might go about explaining her models by asking the right questions, following \citep{belle2020principles}. Few identified tutorials were available for individuals unfamiliar with XAI, those simply seeking an introduction the topic and tools for practical applications \citep{bennetot2021practical, rothman2020hands}. Often tutorials in XAI have been run with a technical research audience in mind such as those recently given at major AI conferences \citep{SamekTUTxai, CamburuTUTxai}, where the purpose was research discovery among peers and not the teaching of fundamentals. These presentations were appropriately technical given the specialist audience. In contrast, a tutorial following the thrust of this report will be presented by the authors at AAMAS 2022\footnote{List of AAMAS 2022 tutorials can be found here: https://aamas2022-conference.auckland.ac.nz/program/tutorials/}. Generally, this course seeks to explore and explain the fundamentals of XAI.

This new course is an online course targeted towards data scientists who are interested in learning about XAI, and navigating this exciting and fast-paced field. Given that, the course is designed to be applicable to industry professionals as well as academic students, such as undergraduates and graduates, who have taken an introductory machine learning course. Note that we include preparatory lectures on machine learning, where we cover key topics ranging from popular models (e.g. random forest, deep learning) to standard training regimes such as cross validation. We provide accompanying code assignments that students can practice with, but we have developed the remaining assignments in such a way that detailed knowledge of machine learning is not necessary. More broadly, the aims of this course align with those individuals whom use machine learning with the purpose of seeking to convince a stakeholder to approve a model. In the interest of simplicity however, we simply state `industry practitioners' with the understanding that they apply to data scientists more broadly.

Consider that machine learning models are increasingly deployed in a wide range of businesses. However, with the increasing prevalence and complexity of methods, business stakeholders may have 
concerns about the drawbacks of models, data-specific biases, and so on. Analogously, data science practitioners are often not aware about approaches emerging from the academic literature, or may struggle to appreciate the differences between different methods, so end up using industry standards such as SHAP. Here, we aim to help industry practitioners understand the field of explainable machine learning better and apply the right tools. 


From the course, we seek to impart a number of key concepts. The first concept taught is that the various explainabilty approaches can be taxonomized such that an industry professional can select a technique by considering explanation types, explanation properties, advantages and disadvantages, and by the model in which they are seeking to explain. The second is that the implementation of various explainability techniques can be done in such a way that a narrative structure is formed which successfully answers potential questions raised by stakeholders. As discussed the narrative structure is inspired by \citep{belle2020principles}, but there is also other recent work providing evidence that explainability techniques are best linked to stakeholder questions \citep{aix360-sept-2019}. The technical details of these approaches will also be covered so that students can have a deeper understanding of the theory behind their implementation. Along with imparting theory, we also aim to provide students experience in applying these explainability techniques using commonly available APIs such that they understand the implementation pipeline (e.g. data cleaning, parameter tuning). Furthermore, we aim to increase student comprehension by raising open questions both during the coding practice sessions as well as for the theoretical discussions. Finally, we want to reflect on areas in the literature of XAI where we may see future research as well as discuss more philosophical concepts such as trustworthiness and transparency.

From an organization viewpoint, after motivating the area broadly, we turn to technical insights, which includes three frameworks: a taxonomic framework provides an overview of explainable ML, and the other two frameworks further study so-called transparent models vs opaque models in that taxonomy. The latter then requires model specific or model agnostic post-hoc explainability approaches, which have their individual limitations and strengths that are also discussed. We also briefly reflect on deep learning models, and conclude with a discussion about future research directions. In terms of exercises, we will focus on 6 popular techniques and try to understand how to use these explainability techniques with coding problems for publicly available datasets.

Intermediate assignments will be evaluated against workbooks that make use of the post-hoc explainability methods discussed in the course. For final assessment, participants can work on a project of their choice, including their own data and demonstrate the ability to mix and match different methods, including at least two not discussed by us, towards a problem statement.  

The codebooks are set up and students are encouraged to use the publicly available Google colab notebooks or similar, as no specific hardware is required \citep{Bisong2019}. The recommended use of Google colab notebooks is due to the reduced hurdle in setting up the coding environment, allowing students fast access to implementing the various XAI techniques. Students simply need to install a file package, containing a dataset and a jupyter notebook covering a specific XAI technique, and upload the contents to their Google Drive in order to quickly begin running the notebook in their web-browser. While it is possible to run these notebooks locally, from our experience we have noted that individual hardware specifications and library dependencies can result in implementation issues.

Individuals will work on their own projects, alone or in groups, and will go through all the stages of the exploration of posthoc explainability, in alignment with the learning outcomes (discussed below); explore data and make initial findings, critique the tools and list shortcomings, list possible future features, and present an analysis. 

The course has 10 lectures, each targeting a set of principles and/or technique, and which are organised as follows. Topics are provided as pre-recorded lectures, and following a flipped classroom model, during the interaction/live sessions there will be a discussion about the content and insights from the pre-recorded lectures. Participants will have an opportunity to debate conceptual aspects during these live sessions. In addition, there will be tutorial sessions that will give participants an  opportunity to engage  in coding exercises and raise practical issues about the corresponding topics.

In broader terms, we believe this course enables inclusivity, empowerment, and responsibility with respect to XAI. In regards to \textit{inclusivity}, the course suggests strategies that can help make machine learning and XAI more accessible as well as help to reach those currently excluded from the rapid development in XAI. As discussed, the current developments in XAI can be very technical and difficult to grasp for those data scientists and industry practitioners not keeping up with the state of the art. By focusing on a small set of techniques and showing that they can be tightly coupled with certain types of questions, provides a reasonable and accessible strategy. On the one hand, it shows data scientists how to better inspect the models that they have built, while also instructing them on the basics of model building. On the other hand, it also allows high level stakeholders, who potentially lack deep technical knowledge, an understanding of what steps need to be taken to approve models, and balance trade-offs between accuracy and transparency. With respect to \textit{empowerment}, the course is designed for computer scientists who have some experience with data, and includes preparatory lectures on machine learning. It gives students the opportunity to engage with machine learning models, debug them, and better relate the performance of the model for their purpose. In terms of enabling \textit{responsibility}, it is widely acknowledged that responsible design in artificial intelligence includes many facets, from bias detection and de-biasing to value alignment. In this broad picture of ensuring that the machine learning model is performing as it should, while accounting for the misspecification of objectives, distribution drifts, and other challenges in model training, explainability is an important ingredient for trusting the model. While there are good arguments to be made about using transparent models in the first place \citep{rudin2019stop}, it is still the case that most businesses and standalone applications will likely be comfortable using conventional machine learning models. In that regard, approaches to explainability and interpretability will play a significant role in checking correctness, debugging, and improving the performance of such models. By focusing on the theory, practice, as well as the narrative structure, we hope to have informed practitioners how to use the techniques in a responsible and careful manner.

To the best of our knowledge, this is the first paper of the sort and so we do not have empirical comparisons  but simply report on structure, content, and learning objectives of the course. We also discuss student feedback which demonstrates positive retention of the course content. We hope this will also be the start of discussion on how to teach this very important dimension to machine learning.

\section{Teaching Aims}


Upon completion of this course we expect students to have learnt to apply XAI techniques such that they can use them to scrutinise and tease apart a given model, consider additional issues such as robustness, determine model fairness, perform model debugging, and improve model accuracy. Learning objectives also include consideration of XAI relevant mathematical details. A deeper discussion of the relevant details can aid students' understanding of the underlying technical details, giving more academically minded students the necessary background to pursue the development of their own XAI implementations. Moreover, a cursory study of the mathematical concepts can help industry professionals understand how the concepts work, but is usually not necessary for the application of the various techniques, at least for most practical assignments.\footnote{Clearly, there are caveats when the theory review is cursory. For example, different implementations of counterfactuals can have different optimisation strategies for how to specify the objective function. These implementations may extract feasible answers, but differing objective functions can also result in diverse answers between the strategies. Therefore understanding the mathematical concepts can become necessary if you want to apply them rigorously.} 


Pedagogical aims of the course as well as the expected learning outcomes are as follows:
\begin{itemize}
    \item Analyze: Describe the context of the machine learning application and why explainability would help, but also scrutinise which kind of explainability technique is necessary.
    \item Design: Define the implementation pipeline for the project; provide a means to clean the data, install and set up one or more post hoc explain ability techniques through a self-chosen set of programming  platforms.
    \item Evaluation: Critically reflect on the results from such techniques and suggest how it helps the problem context. 
    \item Apply: Competently apply a wide range of techniques and tools, also knowing their particular features and drawbacks. Have the foundations to understand new and upcoming methods and techniques. 
\end{itemize}

In particular, because we cover both the theory as well as the practical aspects, on the one hand students would understand the context in which these techniques are deployed but also understand the theory in justifying these techniques. The final project in particular is an opportunity for students to create a narrative and an application and motivate and argue for or against the machine learning model by using a sequence of techniques. In the current installation of the course, the way the final project is designed is to insist on not only using the techniques that students have studied in the class but also consider techniques outside the once discussed during coding exercises. This will mean that they will move beyond the existing teaching material but also understand the difficulties in using less known XAI techniques while appreciating the additional intepretability that these different techniques offer. In particular, some of the lesser known XAI techniques often come with software that is not as thoroughly tested or the documentation is sparse. This might affect the scaleability of the software package as well as the ease of integrating it within an existing pipeline. 
 
On completion of this course, students would learn the following transferable skills:

\begin{itemize}
    \item Problem analysis: analyze the problem related to exploring and communicating data in a specific context
\item Critical thinking: thinking critically about the effectiveness of post hoc explainability for a given challenge, in a given context.
\item Creativity: searching for the right set of explainability solutions to a specific challenge
\item Communication: sensitivity about how to use explainability to explain, improve and debug ML models
\end{itemize}


\section{Course Framework}

The intended audience of the course includes both students and professionals in the area of data analysis and machine learning. We expect a basic understanding of machine learning and/or data analytics, and programming skills in python. A short self-sufficient primer will be given to machine learning, and from there on, the programming exercises will assume some prior understanding of machine learning concepts.

Due to the prerequisites, it should be considered as an introductory Masters-level course, providing foundational skills in terms of an overview of the subject of explainability in machine learning.




\subsection{Objectives}
The objective of the course is to teach the general principles and practice of explainable machine learning. More specifically, the main focus is to enable participants to understand current popular XAI techniques, while also getting an overview of the challenges and potential of post-hoc explainability. Special focus has been put into presenting current techniques as representative instances of particular broader conceptual categories. For example, when presenting SHAP, we first discuss feature relevance explanations in general, and then introduce the specifics of SHAP. This is because XAI is a rapidly developing field, meaning that in a few years there will probably be another technique that will outperform SHAP in popularity. However, the goals and challenges of feature relevance explanations will not change significantly, so interested students and practitioners should be able to follow up future advances in the field. Finally, codebooks and assignments aim at allowing students to figure out, in an interactive way, how such methods can support the understanding and debugging of ML models. 

\subsection{Course structure and content}
In what follows, we outline the structure of the course, which is comprised of $10$ lectures as well as $4$ tutorials over the semester, while the students are expected to submit $2$ assignments and a final project. Each lecture is focused on a specific topic, such as detailing the characteristics, advantages, and disadvantages of a certain XAI technique. In addition, each lecture notes include $3$ open ended questions that aim at initiating a discussion around the topic at hand, which we present in the end of each section.


The purpose of this discussion and following elaboration is to consider the key concepts that have gone into some of the taught XAI techniques. For readers already familiar with these XAI techniques, we suggest skimming the technical definitions. Readers unfamiliar with these technical discussions may additionally benefit from reading the key points, the open-ended questions, as well as the connecting themes that one can conclude by the short exposition.


\subsection{Preliminaries}
\subsubsection{ML preface}
We begin with briefly introducing ML objectives, focusing on the various stages of the general process (i.e. acquiring a dataset, selecting and training a model, evaluating its performance). Following that, we discuss the properties of certain ML models that are going to be used as a reference in the next lectures, in order to demonstrate XAI concepts. Specifically, we make a distinction between transparent models, which include:
\begin{itemize}
    \item Linear/Logistic regression
    \item Decision trees
    \item K nearest neighbours
    \item Rule based learning
    \item Generalized additive models
    \item Bayesian networks
\end{itemize}
and opaque models, such as:
\begin{itemize}
    \item Random forests
    \item Support vector machines
    \item Neural networks
\end{itemize}

The goal of this lecture is to highlight which properties make a model more understandable to a human, emphasising intuition. This serves as a first step towards understanding the necessity of developing XAI techniques in order to enhance opaque models with interpretable features, so they resemble their transparent counterparts.

Open ended questions, such as those below, are presented to students for debate and discussion; such questions are discussed in each lecture:
\begin{itemize}
    \item Can you think of a reason for why opaque models, often, have better performance than transparent models? 
    \item Can you think of cases when opaque models are outperformed by transparent models? (Hint: consider relational data.)
    \item Can you argue about why we may want to solve a ML challenge using both transparent and opaque modelling? 
\end{itemize}

By encouraging students to think about the pros and cons of models and provide use cases, they encounter the challenges when deciding which model to employ between transparent and opaque ones.

\subsubsection{XAI preface}
In this lecture we begin with discussing cases that highlight the need of XAI, such as adversarial examples that drastically alter the model's original outcome, while being indistinguishable to humans. Following that, we talk about the various transparency levels that a model can satisfy \citep{ARRIETAxai}:
\begin{itemize}
    \item \textbf{Simulatability:} A model’s ability to be simulated by a human.
    \item \textbf{Decomposability:} The ability to break down a model into parts (input, parameters and computations) and then explain these parts. 
    \item \textbf{Algorithmic transparency:} The ability to understand the procedure the model goes through in order to generate its output.
\end{itemize}

We emphasize the different requirements of each level, as well as cases where a model could satisfy all of them (such as simple linear models), or none of them (such as multi-layer neural networks). This leads to the observation that, in the latter case, post-hoc methods for inspecting the internal mechanisms of a model are essential to employ them in critical applications. In general, XAI is important in identifying issues such as:

\begin{itemize}
    \item \textbf{Wrong Objective:} A high model accuracy can be flawed if the metric by which the performance is measured is flawed or incomplete in some manner. It could be the issue that the objective was misinterpreted by the model, but having an intelligible system tells a user if a model’s reasoning is correct along with the actual model decision.
    \item \textbf{Inadequate Features:} If certain features are considered inadequate for a final model decision and they are removed from training, it still may be possible to train a model on features which correlate with those that were removed, thus indirectly encoding them. Interpretable models can potentially identify such correlations between features.
    \item \textbf{Training Distribution Bias:} Models may not perform as successfully when they are eventually deployed into the real world. This is evident when the dataset distributions used for model training are different from the input distributions encountered after deployment. It should also be noted that in the real world, distributions can change over time. Including intelligibility can aid in recognizing whether a model is failing to generalize.
    \item \textbf{User Control:} ML systems frequently adjust behavior based on user input. A standard example is seen with online news feeds which show specific stories based on what it predicts as most interesting for a user. Having a user understand ML decision making, can lead to better instructions for the model by the user, if the model performs an undesired action.
    \item \textbf{User Acceptance:} By providing explanations, users are more likely be satisfied and accept a ML decision.
    \item \textbf{Improving Human Insight:} Beyond just using ML to perform automation tasks, scientists can use ML for research purposes with respect to big data. An intelligible model can provide information to scientists based on the data being modelled.
    \item \textbf{Legal Imperatives:} Using ML to assess legal liability is a growing issue, as auditing situations to determine liability requires clear explanations from a model's decision. The European Union’s GDPR legislation decrees citizens’ right to an explanation further strengthens the need for intelligible models.
\end{itemize}

After the importance of XAI is established, the next step is to provide an overview of the ways that research has considered for producing explanations. A first distinction is between global explanations, that explain the model as a whole, and local ones, that attempt to explain the model's prediction for the specific datapoint of interest. In addition, another distinction comes based on the applicability of each technique, where model-agnostic ones are generally applicable to any model, while model-specific techniques are designed to be applied in certain classes of models. Finally, we discuss the various forms of explanation forms that can be found in the literature \citep{ARRIETAxai}:
\begin{itemize}
    \item \textbf{Text explanations:} Produce explainable representations utilizing symbols, such as natural language text. Other cases include propositional symbols that explain the model’s behaviour by defining abstract concepts that capture high level processes.
    \item \textbf{Visual explanation:} Aim at generating visualizations that facilitate the understanding of a model. Although there are some inherit challenges (such as our inability to grasp more than three dimensions), the developed approaches can help in gaining insights about the decision boundary or the way features interact with each other.
    \item \textbf{Explanations by example:} Extract representative instances from the training dataset in order to demonstrate how the model operates. For an example to make sense, the training data has to be in a form that is comprehensible by humans, such as images, since arbitrary vectors with hundreds variables may contain information that is difficult to uncover.
    \item \textbf{Explanations by simplification:} Refer to the techniques that approximate an opaque model using a simpler one, which is easier to interpret. The main challenge comes from the fact that the simple model has to be flexible enough so it can approximate the complex model accurately. In most cases, this is measured by comparing the accuracy (for classification problems) of these two models.
    \item \textbf{Feature relevance explanations:} Attempt to explain a model’s decision by quantifying the influence of each input variable. This results in a ranking of importance scores, where higher scores mean that the corresponding variable was more important for the model.
\end{itemize}

Open ended questions to present at the end of the lecture include:
\begin{itemize}
    \item Can you argue how the XAI evaluation criteria differ or not from the criteria for a ``good'' ML model?
    \item If a medical system offers $98\%$ accuracy over a transparent model that only offers $88\%$ accuracy, what might you prefer, and why?
    \item Would an ensemble of different transparent models be considered transparent?
\end{itemize}

Again, these questions can motivate discussion about timely topics in the deployment of ML, as well as help students appreciate the difference between standard performance measures and explainability. For example, the second one has led to a huge debate on Twitter, which we point students to.\footnote{The thread can be found here: https://twitter.com/geoffreyhinton/status/1230592238490615816}

\subsection{XAI techniques}
Here we propose an indicative set of XAI techniques that could be taught in the course. We begin with  three techniques we consider to be essential (SHAP, counterfactuals, and InTrees), each of them for its own reasons. SHAP is arguably the most popular XAI approach, having well founded theoretical properties, so we believe it is important for students to be familiar with it. On the other hand, counterfactuals bring together philosophy and XAI, approaching the problem from a completely different angle, while also serve as a stepping stone to more advanced causal inference concepts. Finally, InTrees is a model-specific technique that clearly demonstrates the advantages of utilizing them over model-agnostic ones, while it is also clear and simple enough for students to understand.

In addition to the above, we briefly discuss other techniques that could be taught along the three main ones. Their selection was determined by our narrative and how we believe an introduction to XAI should be approached, based on \citep{belle2020principles}. This is why we complement the aforementioned techniques with Anchors, which produces simple propositional rules, visualizations, which are especially valuable when communicating explanations to non-technical audience, as well as deletion diagnostics, which considers the model as a function of the training dataset. Each of these techniques brings a different perspective on XAI, which we think is necessary to form a complete picture of the underlying model's reasoning. A different narrative might motivate more global techniques, deep learning techniques, or symbolic approaches. Moreover, even with the same narrative, we could replace certain techniques with similar ones, for example anchors with LIME. This means there is a lot of flexibility, depending on the students, the goals of the course, and/or the preferences of the instructor. Having said that, Anchors, visualizations, and deletion diagnostics are the techniques we included when delivering the course, so we will present some of the details of the context and style of the lectures, since they probably played a role in the students' answers to the questionnaire they completed at the end of the course. 

\subsubsection{SHAP}
SHAP (Shapley Additive exPlanations) \citep{shap} is a model agnostic value estimation method for explaining individual predications. SHAP learns local explanations by utilising ideas from game theory, so called Shapley Values \citep{shapley}, in order to measure feature attributions. The objective is to build a linear model around the instance to be explained and then interpret the coefficients as the feature’s importance.

Shapley values provide a means to attribute rewards to agents conditioned on the agent's total contribution to the final reward. In a cooperative setting, agents collaborate in a coalition and are rewarded with respect to their individual contribution. In order to apply this technique to ML models, it is necessary to make adjustments so the problem is expressed in a game theoretic manner:
\begin{itemize}
    \item \textit{Setting/Game}: SHAP interprets the game setting as the model prediction on a single input.
    \item \textit{Reward/Gain}: Measured by taking the model prediction on the input x and subtracting the marginal predictions for all remaining inputs.
    \item \textit{Agents/Players}: Are represented by the various collection of feature values of the single instance x that contribute to the output (receive reward). A player can be interpreted as
individual feature values or as a group of feature values, as in the case with tabular data.
\end{itemize}

Having made these adjustments, the Shapley value of feature $i$ equals:
\begin{align*}
    \phi_i = \sum_{S \subset F\setminus \{i\}}\frac{|S|!(|F|-|S|-1)!}{|F|!}[f_{S\cup\{i\}}(x_{S\cup \{i\}})- f_S(x_S)]
\end{align*}

where F is the set of all features, $f_{S\cup\{i\}}(x_{S\cup \{i\}})$ is the models decision when the features in $S\cup\{i\}$ are given as input, and $f_S(x_S)$ is the decision when only features in $S$ are given.

From a pedagogical perspective, introducing SHAP comes with a number of benefits, such as:
\begin{itemize}
    \item It exemplifies how well established mathematical ideas can be adjusted to take on new problems, demonstrating the multidisciplinary nature of ML related research.
    \item Another benefit is that Shapley values are known to satisfy some important properties (local accuracy, missingness, consistency), allowing for a discussion focused on why these properties are important when generating explanations, or why it is important to have such theoretical guarantees.
    \item A recent line of research has explored ways to design models that are ``immune'' to Shapley values explanations, in the sense that although they might be heavily biased, SHAP fails to uncover this behaviour \citep{646264}. Raising this issue can motivate a discussion about possible pitfalls of XAI techniques, as well as the need to examine a model from various perspectives, instead of utilizing only a single technique. 
    \item The current implementation of the SHAP python module comes with an array of different visualisations, which the students can inspect in order to strengthen their understanding of SHAP.
    \item Different choices in how to define $f_S(x_S)$ lead to different variants of SHAP \citep{shap}, which, again, can motivate a discussion about the impact as well as the consequences of each choice, emphasizing the need not to treat XAI techniques as black box solutions, but rather making informed choices based on the application at hand.
\end{itemize}

Open ended questions include:
\begin{itemize}
    \item What are the trade offs and differences between TreeSHAP and KernelSHAP?
    \item How could a biased trained model ``trick'' SHAP by hiding its bias, i.e. assign Shapley values to protected features that do not match their actual importance in the model’s decision?
    \item With KernelSHAP, the sampling for missing values assumes feature independence, is
there a way to remedy this issue? Can you think of possible solutions?
\end{itemize}

These questions aim at motivating a discussion around the assumptions and computational aspects of evaluating SHAP values. They require the students to think about the differences between sampling from the marginal and the conditional distributions, as well as the computational complexity they induce. This question has also been considered in recent academic works \citep{osti}, so interested students can also look into them, as well as alternative views on which other strategies can be employed to compute Shapley values in general \citep{10.1016/j.cor.2008.04.004}.

\subsubsection{Counterfactual}

A counterfactual explanation is a statement that identifies how a given prediction would need to change for an alternate outcome to occur. Key to counterfactuals is the idea of ``the closest possible world'' which signifies the smallest possible change required of a set of variables for the alternative outcome to be predicted \citep{Lewis1973-LEWC}. Scenarios in which this is useful include flipping a model's classification outcome, or reach a certain likelihood threshold.

Counterfactual explanations have historically been associated with philosophy, psychology, and the social sciences \citep{Ruben1990-RUBEE-3}. Furthermore, they inherently convey a notion of closeness, which facilitates the identification of the important factors that lead to a decision. For example, if a loan application is rejected, then providing a counterfactual (i.e. a successful application which is as similar as possible to the original one), makes it easier for a person to identify the important information that is relevant to their specific application. In a sense, counterfactuals highlight why a decision was not made, in contrast to other approaches that aim at explaining why a decision was made.

One of the most popular frameworks for generating counterfactuals for ML models is based on \citep{counterfactual}, where the authors express the problem as:
\begin{align*}
    &\min_x d(x,x_i) \text{   s.t.}\\
    &f(x)=Y
\end{align*}
where $d$ is a distance function, $x_i$ is the factual datapoint, $x$ is the counterfactual one, $f(\cdot)$ is the ML model, and $Y$ is the category we would like the counterfactual point to be classified into. For differentiable models, this problem can be solved using Lagrange multipliers, along with an optimization scheme, such as ADAM \citep{DBLP:journals/corr/KingmaB14}. 

Subsequent works have considered imposing additional conditions in order to generate more informative counterfactuals. For example, \citep{10.1007/978-3-030-58112-1_31} argue that counterfactuals should:
\begin{itemize}
    \item Result in a prediction as close as possible to the desired outcome.
    \item Be as close in value to the original instance as possible.
    \item Minimize the number of features changed.
    \item Use statistically likely feature values.
\end{itemize}
They then proceed to form an objective function that takes into account all these observations to generate counterfactuals.

Introducing counterfactuals is beneficial from a pedagogical perspective since:
\begin{itemize}
    \item They provide an entry point for drawing connections with concepts from causal inference (CI). CI is expected to be one of the most promising future research directions, but it is often challenging for students to grasp the underlying concepts. However, the notion of counterfactuals used in XAI is a simplified version of the ones in CI, so it possible to build on them in order to facilitate the understanding of more advanced ideas.
    \item In addition, discussing the progression from the initial work in \citep{counterfactual} to more recent advances demonstrates how XAI is a dynamic field, where a technique can be refined by taking into account new requirements or desiderata. This could help develop students' critical thinking, enabling them to identify the reasons behind such progressions happen.
    \item Finally, counterfactuals showcase the interplay between XAI and other domains, such as fairness in AI, or applications, like model debugging, all of which exemplifies the interdisciplinarity of XAI related research. For example, by probing a model through generating multiple counterfactuals, we can examine whether changes on sensitive attributes (such as gender) may lead to the model producing a different outcome. If this is the case, then this is a clear indication that the model exhibits biased behaviour.   
\end{itemize}

Open ended questions include:
\begin{itemize}
    \item Can you get multiple counterfactuals for a given instance? If yes, how should we interpret them?
    \item How can we handle discrete features?
    \item How would counterfactuals work with image data?
\end{itemize}

These questions aim at motivating a discussion about the interpretation of counterfactuals, as well as more conceptual issues. For example, the last question prompts the students to think whether it makes sense to consider single pixel perturbations within an image as a meaningful way to uncover information. Informing a person that the model's decision can be changed if a small number of pixels take on a different value is probably not helpful, so, instead, it would be more meaningful to look for counterfactuals with repsect to objects and shapes within an image, as opposed to pixels.

\subsubsection{InTrees}

InTrees \citep{intrees} are a model-specific XAI method for tree ensembles, which take advantage of the tree architecture to produce interpretable explanations. It can be seen as a pipeline of multiple algorithms which:
\begin{itemize}
    \item Extract rules.
    \item  Measure rules/Rank rules
    \item Prune rules, removing irrelevant or redundant variable-value pairs of a rule.
    \item Select rules, choosing a compact set of relevant rules and dismiss redundant rules.
    \item Summarize rules by taking extracted rules and returning a simplified tree ensemble
\end{itemize}

InTrees can explain tree ensembles trained to solve both classification and regression problems. They can be applied to any tree architecture that contains decision trees that split internal nodes using a Boolean check on a single feature and assign the outcome at the leaf node. This includes random forests, regularized random forests, boosted trees, and other tree ensembles. This is important since tree ensembles are very reliable ML models, but it is often difficult to understand due to their complexity, or deploy due to their size, so applying InTrees can result in much more compact models.

One of the advantages of this technique is that it is based on intuitive and easy to explain algorithms, although we are not going to get into the details in this work. In turn, InTrees demonstrate how certain black-box architectures may contain pieces of information that can facilitate the model's understanding. It is worth noting that in the core of this technique lies the idea that although a tree ensembles might be opaque, each of its constituents is transparent, so they can be readily inspected.   

The previous observation perfectly captures the utility of model-specific techniques; instead of relying on universal approximations, develop alternatives that take advantage of the specific characteristics of the model at hand. The majority of model-agnostic approaches make a significant assumptions about the underlying model, which are often violated, compromising the quality of the resulting explanations. Consequently, one of the main drives of model-specific explanations is to reduce the number of assumptions, leading to more accurate explanations. Introducing InTrees has the benefit of clearly demonstrating the concept in a simple way, as opposed to more mathematically challenging alternatives, for example neural network LRP explanations \citep{10.1371/journal.pone.0130140}. This should improve the student's understanding of why model-specific explanations are important, without obscuring the message with overly complex technical details.

Open ended questions include:
\begin{itemize}
    \item What is the tradeoff between frequency and error in practical scenarios? Which should we aim to optimize?
    \item Can you argue with examples about what happens if pruning is not applied?
    \item Can a similar idea be applied to non-tree ensembles (e.g. SVM, neural networks)? If so, how do you think this would be possible?
\end{itemize}

This time the questions touch upon both practical issues of deploying InTrees in practice, as well as more conceptual aspects. In particular, the last question can motivate a discussion around the features of tree ensembles that make them suitable for developing techniques such as InTrees. Contrasting them to different models can highlight their dissimilarities, as well as how properties that can be found in these models may lead to new approaches. For example, while neural networks do not have internal splitting rules, their layered structure allows for back-propagating messages that quantify the importance of each feature, such as in LRP \citep{10.1371/journal.pone.0130140}.

\subsubsection{PDP/ICE}

Another prominent means of explaining a ML model is using visualizations, especially when communicating explanations to a non-technical audience. A Partial Dependence Plot (PDP/PD) \citep{friedman2001} plots the average prediction for a feature(s) of interest as the feature’s valuation changes. These plots can reveal the nature of the relationship between the feature and the output, for example, whether it is linear or exponential. PDPs present global explanations, as the method factors all instances and provides an explanation regarding the (marginal) global relationship between a feature and the model prediction. Assuming we are interested in examining the partial dependence of the model $f$ on a feature $s$, we have to compute:
\begin{align*}
    f^*(x_s) = \sum_{i=1}^N f(x_s, \textbf{X}_{-s}^{(i)})
\end{align*}

where $N$ is the cardinality of the dataset, and $\textbf{X}_{-s}^{(i)}$ is the $i-th$ datapoint, excluding feature $s$. We see that the partial dependence function provides the average marginal output for for a given value of $x_s$. Furthermore, it is not difficult to extended this method to account for the partial dependence of a function on more than one features, however, this is usually done for one or two features, due to our inability to perceive more than $3$ spatial dimensions. 

Individual Conditional Expectation (ICE) \citep{goldstein} plots model predictions for multiple instances, where for all instances, only the feature of interest changes value, while the remaining feature values for a given instance are held constant. This plot shows the feature-value and model prediction relationship, for each instance as a separate line, as opposed to the single average predictive line with PDP. Some of the advantages of employing ICE are:
\begin{itemize}
    \item PDPs may hide some of the heterogeneous relationships between feature value interactions by averaging them out. Consider that two opposing, but equally valued, influences on model prediction can be cancelled out when averaged.
    \item PDPs provide a view into the average predictive behavior of a model with respect to a single feature, but the validity of this predictive behaviour is diminished by any possible interactions with the feature being plotted and the remaining features in the data.
    \item ICE plots are able to plot more accurate relationships even with the presence of highly correlated features.
\end{itemize}

Finally, there is an interesting relationship between these two plots, as averaging the ICE plots of each instance of a dataset, yields the corresponding PD plot.

Open ended questions include:
\begin{itemize}
    \item What are the key limitations of PDP/ICE?
    \item Roughly sketch a $3$-dimensional PDP and ICE plot. Based on that, argue about whether this makes explaining the model easier or not.
    \item Instead of averaged out values, how can you show the minimum and maximum for the features in PDP? Is that useful?
\end{itemize}

This time the questions are more focused on the limitations of PDP/ICE, as well as possible ways to manipulate them in order to produce new visualizations. Having a discussion around these topics can help students identify alternative visualizations that remedy some of these issues, for example ALE plots. Furthermore, coming up with ways to modify the existing plots to convey different kind of information can instil confidence into the students that they have grasped the important details of PDP/ICE.

\subsubsection{Anchors}

Rule-based classifiers have been traditionally utilized due to their transparency, since they are easy to inspect and understand. Anchors \citep{AAAI1816982} is a XAI technique that builds on this principle, aiming at generating simple if-then rules to describe a model's reasoning. They explain individual predictions locally by identifying a decision rule that ``anchors'' the prediction in question, thus they operate on instance level. A rule anchoring a prediction implies that changes to the remaining feature
values do not impact the prediction. Anchors do not assume a dataset a priori, and so a perturbation-based strategy together with the black-box model are used to generate local explanations. Rules are reusable if the conditions of the rule are met and are measured by the rules \textit{coverage}, which defines explicitly which instances, specific feature values, the rule applies to. Rules also consider \textit{precision}, which is defined as the fraction of instances in which the rule is correct.

Formally, an anchor, $A$, is defined as the solution of the following problem:
\begin{align*}
    \max_{\text{A s.t. }P(prec(A) \geq \tau) \geq 1-\delta} cov(A)
\end{align*}

where $prec(A)=\mathbb{E}_{D(z|A)}(\textbf{1}_{f(x)=f(z)})$, $D(z|A)$ is the data distribution given $A$, $f(z)$ is the ML model, $\textbf{1}$ is the indicator function, and $cov(A)=\mathbb{E}_{D(z)}(A(z))$. In words, this optimization problem looks for if-then rules where the preconditions (the if-part) contains conditions that are satisfied by as many instances as possible, while requiring that these points also satisfy the then-part. This way the resulting rules are not based on niche characteristics of the specific datapoint at hand, but are as generally applicable as possible.

Open ended questions include:
\begin{itemize}
    \item  What would you choose between anchors with high precision and low coverage vs anchors with low precision and high coverage?
    \item Can you give examples of rules that might apply to recent data you have encountered? Can you argue about what precision/coverage you expect them to have?
    \item Compare anchors to other local explainability techniques, such as SHAP. What are the advantages and disadvantages compared to it?
\end{itemize}

The questions above can motivate a discussion about the utility of anchors under different conditions, as well as a comparison between rules and other forms of explanations, such as importance scores. This gives students the chance to compare different techniques and then potentially reach the conclusion that the answer to this depends largely on the audience the explanation is intended for. This is an important topic that has gained a lot of traction within the academic community, so students can familiarize themselves with timely topics, as well as deepen their understanding on communicating explanations \citep{10.1145/3351095.3375624}.

\subsubsection{Deletion Diagnostics}

Deletion diagnostics is a technique which investigates the model as a function of its training data. It considers the measurable impact on model prediction (or the decision boundary) that removing a particular instance from training has \citep{doi:10.1080/00401706.1977.10489493}. By removing an instance with significant influence from training, deletion diagnostics can help with model debugging and comprehending model behavior. In this context, an instance is considered to be influential if its removal causes the parameters of the trained model to change significantly or results in notably different predictions on the remaining instances. This can aid in debugging by locating influential instances in the training set that are contributing errors on the test set. Locating influential instances can also help with revealing information regarding what features the model relies on for predictions.

Furthermore, this task is related to counterfactuals, too, since it answers the question ``how would the parameters of a trained model change or its prediction change in the case of retraining the model on a training set with a given instance removed?''. This provides for an interesting observation which could motivate a discussion regarding additional ways for counterfactuals to be utilized in XAI. 

Identifying influential instances is important since they invert the relationship between model and data where we now look at the model as a function of the training set. Influential instances can inform how specific feature values influence model behaviour, they can also be used to identify adversarial attacks, they can help in debugging by identifying instances which result in model errors, and they can help in fixing mislabelled data \citep{10.5555/3305381.3305576}. All of these are significant information when explaining a model, which could also find additional applications, for example reducing the size of the training dataset. This could be achieved by retaining only the influential instances an then retraining the model using them, since these instances can express model behaviour where it is most sensitive, contributing to a better understanding of model behaviour.

When it comes to measuring a feature's influence, various measures have been proposed, such as Cook's distance:
\begin{align*}
    D_i = \frac{\sum_{j=1}^N (\hat{\textbf{Y}}-\hat{\textbf{Y}}_{-i})}{pMSE}
\end{align*}
where $D_i$ is the influence of the $i$ datapoint, $\hat{\textbf{Y}}$ is the prediction of the model trained on the full dataset, $\hat{\textbf{Y}}_{-i}$ is the prediction of the model trained on the same dataset, but excluding instance $i$, $p$ is the number of features, and $MSE$ is the mean square error. Looking at the formula we see that Cook’s distance is a measure of the aggregate influence of the $i$ instance on all parameters of a trained model, but alternative measures exist, too.

Influence functions \citep{influence} are an alternative to standard methods of deletion diagnostics, wherein the removal of an instance $i$ is approximated, and the model does not need to be retrained with instance $i$ removed. This makes it more efficient to estimate the influence of a datapoint on the final model, since, arguably, it is very computation intensive to retrain the model each time want to perform this kind of analysis.

Open ended questions include:
\begin{itemize}
    \item Can you explain how deletion diagnostics can be done efficiently without retraining?
    \item Do you think deletion diagnostics can be applied to random forests?
    \item What do you think is the relative usefulness of deletion diagnostics compared to influence functions?
\end{itemize}

This time the questions are mostly on the practical side, aiming at motivating a discussion around the differences of deletion diagnostics and influence functions. This encourages students to think about scenarios where, for example, only one of them is applicable, as well as the advantages and disadvantages of employing each technique.

\subsection{\textbf{Future research directions}}

After introducing all the techniques discussed in the previous section, the remaining lectures reflect on XAI. More precisely, we first talk about the link between certain types of questions and types of explanations, which as a topic is forming an emerging line of research \citep{DBLP:journals/corr/abs-2007-05408,Bhatt2020ExplainableML,10.1145/3313831.3376590}. Finally, we talk about various directions that the XAI might go from here.

\subsubsection{XAI Strategies}

The goal of this lecture is to get into the place of a putative data scientist, say Jane, in order to put together the insights they have gathered throughout the previous lectures. We assume that Jane's work is on building ML models for loan approvals. As a result, she would like to consider things like the likelihood of default given some parameters in a credit decision model. We discuss how she might go about explaining her models by asking the right questions, recommending a simple strategy and sample questions.

After trying out transparent models, without achieving high accuracy, she decides to utilize Random Forests for her application. In order to inspect the internal reasoning of the model, she also employs SHAP in order to generate feature importance scores for her variables. However, when presenting the explanations to stakeholders some interesting questions are raised: Could it be that the model relies heavily on an applicant’s salary, for example, missing other important factors? How would the model perform on instances where applicants have a relatively low salary? For example, assuming that everything else in the current application was held intact, what is the salary’s threshold that differentiates an approved from a rejected application?

Jane knows that these questions cannot be addressed using SHAP, since they refer to how the model’s predictive behaviour, , so she will have to use additional techniques to answer them. As a first step, she decides to employ ICE plots, to inspect the model’s behaviour as a function of salary, holding everything except salary constant, fixed to their observed values. Jane discusses her new insights, however, there is a new issue to address, since, in the test set, there is an application that the model rejects, which comes contrary to what various experts in the bank think should have happened. This leaves the stakeholders in question of why the model decides like that and whether a slightly different application would have been approved by the model.

After giving it some thought, Jane realizes that this is a what-if question, so counterfactual explanations could help her address it. She applies this approach and she finds out that it was the fact that the applicant had missed one payment that led to this outcome, and that had he/she missed none the application would had been accepted. Stakeholders think this is a reasonable answer, but now that they saw how influential the number of missed payments was, they feel that it would be nice to be able to extract some kind of information explaining how the model operates for instances that are similar to the one under consideration, for future reference.

Jane thinks about it and decides that the best way to convey this information to her audience is to use simple propositional rules. To this end, she employs anchors to generate ``if-then'' rules that approximate the opaque model’s behaviour in a local area. The resulting rules would now look something like ``\textit{if salary is greater than 20k £ and there are no missed payment, then the loan is approved}''. 

The stakeholders are happy with both the model’s performance and the degree of explainability. However, upon further inspection, they find out that there are some data points in the training dataset that are too noisy, probably not corresponding to actual data, but rather to instances that were included in the dataset by accident. They turn to Jane, in order to get some insights about how deleting these data points from the training dataset would affect the models behaviour. To account for this issue, Jane decides to go with deletion diagnostics, so she can identify datapoints that could significantly alter the model's performance, if left out. All of these helped the stakeholder understand which training data points were more influential for the model.

Finally, the stakeholders ask Jane if it is possible to have a set of rules describing the model’s behaviour on a global scale, so they can inspect it to find out whether the model has picked up any undesired general pattern. Jane thinks that they should utilize the Random Forest’s structure, which is an ensemble of Decision Trees. This means, that they already consist of a large number of rules, so it makes sense to go for an approach that is able to extract the more robust ones, such as inTrees. 

Having inspected the model through all these angles the stakeholders feel confident in their understanding of its internal reasoning. The point of the analyses is to emphasize that a single technique is often not enough to get an adequate picture, but combining multiple techniques to inspect different aspects of the model should be preferred, instead.

Open ended questions include:
\begin{itemize}
    \item Why are post-hoc methods needed for transparent models?
    \item Can local explanations also be visual explanations?
    \item What is the taxonomic position of SHAP, counterfactuals, PDP, and InTrees?
\end{itemize}

This time the questions serve more as a means to revisit things already discussed in previous lectures, since the Jane example provides for a nice opportunity to take a look at an end-to-end process that combines most of the insights the students have gained so far.

\subsubsection{Future research directions}

The final lecture of the course is about the future of XAI related research. Its goal is to discuss limitation of XAI and prepare students for the next generation of techniques, as well as to provide an overview of which concepts are likely to play a central role in the future. This way the interested students have the chance to study these concepts in advance, so they have the prior knowledge required to potentially grasp future techniques. Of course, it is not possible to be exhaustive and cover all directions, instead we provide an indicative list of current XAI limitations. 

An important point of emphasis for students is to realize that limitations arise both on a technical and a practical level. Most of the existing techniques, especially model-agnostic ones, require resorting to approximations. This means that there is always the danger that the resulting explanations might not be accurate, or even be misleading. Furthermore, existing approaches are not really able to identify spurious correlations and report them back. Due to this, it is possible for features to look like they have a strong influence on each other, when, in reality, they only correlate due to a confounder. A possible resolution to these issues could be introducing more concepts from causal analysis, which is already a major drive in related areas, such as fairness in ML. For example, if an explanation was accompanied by a causal model it would not be difficult to check for any spurious correlations.

On a more practical level, developing XAI pipelines to explain a model it is still an open research question. Currently, there is no consensus regarding neither the characteristics of a good explanation, nor the way of combining existing techniques in order to adequately explain a model. While there is some overlap between the various explanation types, for the most part they appear to
be segmented, each one addressing a different question. This hinders the development of pipelines that aim at automating explanations, or even reaching an agreement on how a complete explanation should look like. On top of that, it is not clear whether explanations should be \textit{selective} (focus on primary causes of the decision making process), or \textit{contrastive} (indicate why a model made decision X, and provide justification for deciding X rather than Y), or both, and how to extract such information from current techniques. Audiences in XAI can include experts in the field, policy-makers, or end users with little ML background, so intelligibility should be varied in its explanations depending on the knowledge level and objectives of the audience. Interdisciplinary research combining psychology, sociology, and cognitive sciences can help XAI in delivering appropriate explanations \citep{Miller2019-MILEIA-6}.

Open ended questions include:
\begin{itemize}
    \item What do you think are the most important limitations of XAI?
    \item Can you suggest ways to automate XAI?
    \item Can you suggest ways to address the potential dangers of transparency?
\end{itemize}

These questions give students the chance to express their views on how to take on these challenges. \footnote{Articles that address related questions can also be found in the literature, for example \citep{DBLP:series/lncs/Weller19}, which interested students are pointed to, so they can study a more in-depth analysis.} This can motivate a discussion around what they think are the most important topics moving forward, as well as how the future of XAI might look like.

\subsection{Assignments}
While the course content includes general information as well as mathematical theory, students assessment is purely practical. As the course is aimed at industry practitioners, a greater emphasis is given to the applying the XAI techniques in a real world setting. Variants of the the course focused on teaching MSc students might emphasise the various algorithmic formula and mathematical derivations for purposes of developing future XAI researchers.

The assignment structure includes four jupyter notebooks \citep{Kluyver2016jupyter} namely SHAP, counterfactuals, Anchors, and InTrees (PDP/ICE and deletion diagnostics omitted for simplicity), each of which demonstrate a singular XAI technique, which provides questions that students must analyse and answer for assessment. Questions include technical code based implementations of a specific XAI technique and short answer questions asking for student interpretation of the outputs. A final project is also assigned which asks students to select a dataset and model of their choice and describe a problem requiring explainability, and consequently implement a minimum of two post-hoc explainability methods. The application of these methods will be used as evidence by students in their discussion of the model's performance on the dataset. Here, the results and the implications from the various XAI techniques will used by the students to convince a stakeholder to either accept or reject a model.

\subsubsection{Workbooks}
The structure of the workbooks is repeated for all four XAI techniques. As stated, the use of python jupyter notebooks allows for ease in implementation by students \citep{Kluyver2016jupyter}. Specifically, when run using Google colab \citep{Bisong2019}, the majority of workspace setup/imports is handled automatically. Within each workbooks is a number of text cells which contain relevant information as well as summary questions. These summary questions, which are presented after a technical process, introduced in the code, are designed to help students review what was just learned. 

The flow of each workbook is as follows. At the beginning of each workbook, students will first import the relevant python libraries (\texttt{pandas}, \texttt{numpy}, etc.) as well as the library associated with the XAI technique which, in general, is not preinstalled on Google colab and so we will directly install the package (i.e \texttt{!pip install alibi}). Following the initial setup, we then being basic machine learning practices such as \textit{Data preprocessing}, where we visually check our data, assess and treat potential missingness, potentially create new labels
by feature engineering new attributes, and
dropping irrelevant columns.


The next step is \textit{Modelling}, here we train our model, assess performance, and potentially perform hyper-parameter tuning to optimise the model prior to applying a specific XAI technique. Each XAI codebook uses an arbitrary model, majority of which are black-box models, and performance is measured by looking at the accuracy, precision, recall, and f1 scores. 

The third section of the workbook is where we apply one of the XAI techniques. This section will be specific to each individual XAI technique but in each case we will apply the technique on the model trained in the \textit{Modelling} step. In general, the XAI technique will be applied in a few cases (singular instances for local explanations or various output formats) and summary questions will be present to test the student's knowledge in the techniques application. Each workbook then ends with a series of assignment questions which, if the assignment is to be assessed, will comprise a portion of the submission work. Assignment questions will include 2-3 short answer questions regarding the technique and/or potential concerns in applying a given model, followed by 3-4 technical questions where the student will need to apply the XAI technique either on another model or on an augmented dataset. Again, the assignment and summary questions are specific to each XAI technique. We cover aspects of the workbooks for SHAP to demonstrate the potential formulation for other methods.

\begin{figure}[ht]
\caption{Demonstration of applying TreeSHAP in the corresponding SHAP workbook.}
\centering
\includegraphics[width=0.8\textwidth]{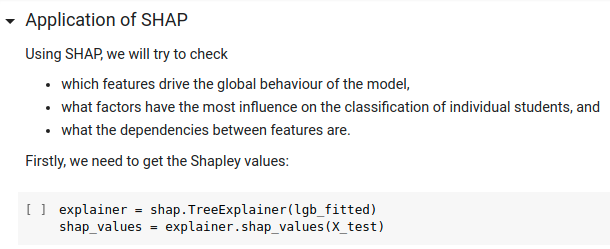}\label{fig 1: shap}
\end{figure}

\paragraph*{SHAP: } As discussed, the flow of each workbook is consistent for each technique (data preprocessing, modelling, XAI technique), with the implementation of the specific XAI technique being distinct. In the case of SHAP, the workbook first imports the relevant library (\texttt{!pip install shap}) and after preprocessing the data and tuning the black-box model, for the workbook an LGBM model, we then fit a TreeSHAP explainer using the model and the dataset, see figure \ref{fig 1: shap}. As LGBM is a tree ensemble method, we take advantage of the model-specific SHAP implementation. Following code cells in the workbook implement various SHAP plots (e.g. \texttt{shap.summary\_plot(shap\_values[0], X\_test)}, \texttt{shap.decision\_plot(explainer.expected\_value[1], shap\_values[1][20], X\_test.iloc[20,:])}), following which we include a text cell discussing the previous output.

Included throughout the workbook are the summary questions, which are designed as general discussions points following the implementation of code in some of the cells or after the introduction of a machine learning or XAI technique. Questions (1-2) are specific to assessing the black-box model and interests of the stakeholder, while questions (3-10) are questions specific to SHAP and its implementation. In general, these questions will be discussed openely with students during tutorial sessions.

\begin{enumerate}
    \item How would you interpret $50\%$ precision in the table above?
    \item Which metric from the above do you think would be of the most interest to a stakeholder interested in a model that aims to predict students' performance?
             \begin{itemize}
         \item \textit{This summary question brings attention to one of the learning aims, that is these techniques can be used as apart of a greater explanatory narrative.}
      \end{itemize}
    \item Generate the same plots for instances in class 0. What is the link between these graphs and what has been generated for class 1?
    \item What are the 5 most important features which are driving this model's decisions?
             \begin{itemize}
         \item \textit{Here students are asked to read the output of one of the SHAP plots.}
      \end{itemize}
    \item How would you interpret the horizontal axis of the first summary plot above? What does a SHAP value of $-1.5$ mean?
        \begin{itemize}
         \item \textit{Students are asked to discuss the features of a SHAP plot as well as SHAP values. As there is a bit of a learning curve to understanding SHAP plots, these discussions are important to help students understand the outputs.}
      \end{itemize}
    \item Can we express SHAP values in terms of probability? Justify your answer.
         \begin{itemize}
         \item \textit{A more open-ended question asking students to reflect how SHAP values can be interpreted.}
      \end{itemize}
    \item According to these plots, which student is most likely to fail (assuming our model is appropriate)?
         \begin{itemize}
         \item \textit{Students are asked to analyse the SHAP outputs for a number of local explanations.}
      \end{itemize}
    \item Are there any features which are important from the perspective of predictions on the local level for these 3 students and which could indicate some fairness issues?
         \begin{itemize}
         \item \textit{This question brings the notion of fairness to the students' attention as well as a larger discussion of local explanations.}
      \end{itemize}
    \item Produce a local plot as above for another student (feel free to select your own observation). How different is it from the global picture?
      \begin{itemize}
         \item \textit{Here students can apply a SHAP plot on another instance. Getting them more familiar with the SHAP methods.}
      \end{itemize}
    \item What does the horizontal spectrum on the top of the graph show? What do those values mean?
         \begin{itemize}
         \item \textit{Students are asked to discuss the features of a SHAP plot.}
      \end{itemize}
\end{enumerate}

At the end of the SHAP workbook we provide students assignment questions which are to be submitted for assessment. These questions will require the students to implement the techniques previously demonstrated in the workbook as well as require investigations online for methods not shown. Overall, questions can be separated into technical (coding focused) and short answer. For the SHAP workbook we classify questions (1-5) as being primarily technical and questions (6-8) as short answer, requiring student interpretation.
\begin{enumerate}
    \item Use the public dataset introduced in this tutorial and apply an XGBoost model. Your outcome variable will be Portuguese language scores pooled into class 0 and 1 in the same way as in this notebook (feel free to skip any hyperparameter tuning operations). Make predictions on your test set and produce a set of measures that describe the model's performance.
     \begin{itemize}
         \item \textit{Students are asked to apply a new model, different from the tutorial section of the workbook, and simply assess its performance.}
      \end{itemize}
    \item Using SHAP summary plots, what are the 5 most important features in the model?
         \begin{itemize}
         \item \textit{Here students are asked to apply SHAP and assess which features are the most significant to the model.}
      \end{itemize}
    \item Create a decision plot for all observations and all features in your test set, highlight misclassified observations and create decision plots for the set of misclassified observations and for 4 single misclassified observations. Then include force plots for all observations as well as for the set of misclassified observations.
        \begin{itemize}
         \item \textit{This questions requires students to identify incorrect predictions of the model and then produce a series of SHAP plots for these points. The purpose is to see if any feature-value pairs have a significant correlation with inaccurate predictions. }
      \end{itemize}
    \item Make SHAP dependence plots of the 4 most important features. Use sex as the feature possibly influencing SHAP outputs. This done by setting the \texttt{interaction\_index} as ``sex".
        \begin{itemize}
         \item \textit{Students are now asked to use SHAP plots to see if there is any bias in the model.}
      \end{itemize}    
    \item In the light of the plots from 3 and 4, discuss whether the interaction effect between sex and other features can meaningfully impact decisions of your model.
            \begin{itemize}
         \item \textit{The first short answer questions asked students to interpret the outputs from the previous questions. The aim is for the students to list a series of observations based on the generated SHAP plots. }
      \end{itemize}
    \item Discuss how various SHAP-based graphs can be used in the process of model validation.
            \begin{itemize}
         \item \textit{A more open-ended short answer question, asking students to discuss the various SHAP graphs and their uses. Evidence from the previous SHAP plots would support the student's response for this question.}
      \end{itemize}
    \item Write a paragraph for a non-technical audience explaining how your model makes decisions based on SHAP outputs. Ensure the text is clear of jargon!
       \begin{itemize}
         \item \textit{Here students are expected to write a longer short-answer response, one that asks students to summarise the previous SHAP plots and produce a formal analysis on what can be inferred. Importantly, students should tailor their response to a non-expert. }
      \end{itemize}
\end{enumerate}

\subsubsection{XAI Final Project}


The purpose of the final project is for students take what they have learned and apply it in a practical setting. The final project will ask that students analyse a machine learning problem and motivate the corresponding context, design the pipeline, apply the appropriate XAI techniques, and finally evaluate the results and argue how the overall problem was addressed. The minimum time commitment expected from students is 14 hours.

The final project will require selection of a dataset and a black-box model (or models) in order to build a case scenario for a stakeholder. For example, \textit{``Taking a credit scoring dataset, and the XGBoost model, convince a banking institution to reject the model using (at least) technique 1 and technique 2''}. Students will use XAI to either argue for a model or against a model, and the idea is to convince a stakeholder to implement or dismiss a model based on what the student was able to discover from applying XAI techniques. The goal with the project is that students will use the various XAI techniques to construct a narrative using their interpreted explanations to support their argument.

By design, the project is open-ended, but students will be required to submit a notebook with their findings as well as a written summary detailing their case. The final project will ask that students be able to implement multiple XAI techniques, at least a couple not considered in class, and support their case scenario using points learned in the lectures.

\section{Evaluation Methodology}
{To assess the effectiveness of the course, we used a modified version of the Course Experience Questionnaire (CEQ) and a general performance analysis based on the coding assignment. Sixteen survey responses were collected from the students who attended the XAI Course through the University programme} and at the time of writing an instance of this course was deployed with an industry partner. We are still collecting responses on this questionnaire from the partner and hope to include an analysis in an extended report. Eight of the respondents were postgraduate research students, 5 were doing non-academic work, and 3 responded ‘Other’. Prior to taking this course, $87.5\%$ of respondents were not familiar with the topic of XAI. Most of the respondents had at least some practical experience of working with ML ($81.3\%$) and theoretical knowledge of ML ($81.3\%$). One respondent had no theoretical or practical knowledge of ML, and two respondents had some theoretical, but no practical ML experience. Four respondents rated their practical experiences in ML as close to expert level, and six rated their theoretical ML knowledge close to expert level.

{Conventionally the CEQ is employed as a measure of teaching quality of university courses \citep{ramsden1991performance}. The questionnaire assumes a strong connection between the quality of student learning and student perceptions of teaching \citep{mcinnis1997defining}. When combined with the additional course assessment items and adapted to the course context it can reliably be applied as a domain-neutral indicator of university course quality \citep{griffin2003development}. For this study 12 items were selected from the original version of the CEQ based on their relevance to the online XAI course context and purpose of this study (See Appendix). These items were scored on a 5-point Likert-type rating scale from ‘strongly agree’ to ‘strongly disagree’. In the modified questionnaire we included items from the scales of Good Teaching (e.g., the teaching staff gave helpful feedback), Appropriate Workload (e.g., I was generally given enough time to understand the things we had to learn), Clear Goals and Standards (e.g., it was clear what was expected of me in this course), the Appropriateness and Effectiveness of Sources of Information and Course Materials (e.g., the materials needed to do the course were readily available), and Overall Satisfaction. In addition to the CEQ items, we used context-specific items to gather information about (i) pre-existing skills and theoretical knowledge of XAI; (ii) pre-existing skills and theoretical knowledge of Machine Learning (ML); (iii) satisfaction of the diversity of the taught techniques; (iv) course ability to build understanding of XAI techniques; (v) level of comprehension of the conceptual distinctions, advantages, and disadvantages of the XAI techniques covered in the course; (vi) success in meeting the four pre-set learning objectives. These responses were also based on a five-point Likert scale. Finally, in an open-ended manner, students were asked to list their favourite aspect of the course and to suggest anything that could help to improve the course in future.}  

{To support the self-reported assessment of the course, we also analysed the general students’ performance. We examined 19 students’ performance data of the first course assessment (SHAP codebook) and observed the trends, such as aspects of the assignment that students were struggling with the most as well as the aspects students had additional questions about. This way we sought a more objective assessment of course effectiveness and limitations.}

\subsection{Results}
{The results were divided into three parts. Overall CEQ scoring of the responses, quantitative and qualitative analysis of the individual statement ratings and qualitative analysis of the SHAP assignment results. The CEQ (Q10) raw scores were recorded as follows: a raw score of 1 (`strongly disagree') is recoded to $-100$, $2$ to $-50$, $3$ to zero, $4$ to $50$, and $5$ (`strongly agree') to 100, eliminating the need for decimal points. The scoring of negatively worded items was reversed. In interpreting CEQ results, a negative value corresponds to disagreement with the questionnaire item and a positive value to agreement with the item. Positive high scores indicate high course quality as perceived by graduates. The responses revealed a high positive overall score of $12,050$, which indicated high course quality as perceived by respondents. The CEQ is widely accepted as a reliable, verifiable, and useful measure of the perceived course quality} \citep{griffin2003development}.

Analysis of the individual ratings of the statements 6-9, further showed the high ratings of the course quality as perceived by respondents. For the statement (Q6) ``please rate how confident do you feel in applying the XAI techniques you learned in your own models from a score 1 (not at all confident) to 5 (very confident)'', the average rating was 3.5 with $50\%$ of respondents rating their confidence as either 5 (very confident) or 4 (confident). For the statement (Q7) ``please rate how satisfied are you from the diversity of XAI techniques covered in the course from a score 1 (not at all satisfied) to 5 (very satisfied)'', the average rating was 4.44 with $87.5\%$ of respondents rating their satisfactions as either 5 (very satisfied) or 4 (satisfied). For the statement (Q8) ``please rate how much do you feel your understanding of XAI was benefited by the course from a score 1 (not at all) to 5 (very)'', the average rating was 4.56 with $87.5\%$ of respondents rating the benefits to their understanding as either 5 or 4. For the statement (Q9) ``please rate how much do you feel you have comprehended the conceptual distinctions, advantages, and disadvantages of the XAI techniques covered in the course from a score of 1 (not at all) to 5 (very)'', the average response was 4.19, with $87.5\%$ of respondents rating their comprehension as either 5 or 4. High ratings on each item indicate high perceived course effectiveness in fostering comprehension of the separate XAI aspects, building overall understanding of XAI techniques, as well as indicating a satisfying amount of diversity of XAI techniques covered in the course. Lower rating of the Q6 suggests that the course might benefit from including more practical aspects that would support students' ability to apply taught XAI techniques in their own ML models.

For the statement (Q11) ``please rate how successful you were in meeting the learning objectives from a score of 1 (not all successful) to 5 (very successful)'', $37.5\%$ of respondents reported being `very successful' and $56.25\%$ `successful' in meeting Analyse learning objective; $25\%$ of respondents were `very successful' and $50\%$ `successful' in meeting Design objective. Students reflected that they were less successful in meeting the Apply and Evaluation learning objectives. For the Apply item $20\%$ of respondents said they were `very successful', $60\%$ successful, $13.3\%$ `somewhat successful' and $6.67\%$ `neither successful nor unsuccessful'. For the Evaluation item, $18.75\%$ respondents reported that they were `very successful', $62.5\%$ that they were `successful', and $12.5\%$ `somewhat successful' and $6.25\%$ `neither successful nor unsuccessful'.

The qualitative analysis of the open-ended questions revealed specific aspects of the course that were recognised as respondents' favourite. Eleven respondents answered the question (Q12) `what was your favourite aspect of the course?'. Workbooks were mentioned by five respondents. Students appreciated being able to try out the theoretical course aspects in a practical way using the provided workbooks. For example, S-5 said: \textit{``trying out the techniques in the workbooks''}. S-9 said: \textit{``practical application in lab books''}. S-7 said: \textit{``the workbooks and assignment questions. They had the right mix of theory and practical aspects''}. Five respondents mentioned recorded lectures and tutorials. Students appreciated being able to discuss the course material in the online tutorials and lectures. S-1 put it: \textit{``the discussion and Knowledge exchange during Lectures and Tutorial classes''}. S-2 said: \textit{``the meaningful discussions and open-ended questions''}. S-8 said: \textit{``the ambience of the teams' sessions is done with a ``brainstorming'' approach which gives us the opportunity to discuss ideas, bring questions from the real world and hear different opinions''}. Respondents also mentioned open-ended questions, sufficient examples, and assignment questions. Overall responses were very positive, for example S-2 said: \textit{``I was very impressed with the structure and delivery of the material...It [the course] made me not only appreciate the XAI fundamentals but the whole approach towards applying ML algorithms...I consider myself very lucky for selecting this course and I believe it has helped me tremendously in my understanding of ML projects.''} S-6 said: \textit{``I have understood why the area of XAI techniques has gotten attention and is important to make AI / ML available for general use in the Data Analytics Project.''} S-8 reflected: \textit{``The topic of XAI it's very interesting! Thank you for including it in the program and giving us exposure to these approaches.''}

Six respondents made suggestions for improving the course in future (Q13). Two suggested that more participation from other students would be beneficial, potentially scheduling the online classes to accommodate those who are working normal working hours. Other suggestions were to state in the course description that prior ML knowledge is necessary and provide examples of opaque models like neural networks and image recognition.
To provide objective assessment of the course effectiveness we have analysed the responses of the SHAP codebook assignments and observed several performance trends in the students' submissions and questions of the class forum. Around 30 of student questions regarding the assignment were theory-based questions. Students wanted explanations regarding some of the technical aspects of the algorithm or more in-depth explanations regarding scoring metrics. Most of the questions regarding coding aspects of the assignment were concerned with interpreting the questions. Often it seems the verbiage of the questions was not immediately clear, so students would frequently message for further clarification. The analysis of the assignment results revealed that students were able to execute the code but encountered problems in the analysis of the results. For example, students' performance was poorer when they were asked to compare feature importance and interpret/demonstrate understanding of the outputs, for example with various SHAP graphs. Further issues with coding happened usually due to the library updates, students did have technical issues, but it was often because the XAI package changed resulting in different output or requiring different formatting of the data. This meant that we had to prepare multiple versions of the codebook. Overall, $100\%$ of students passed the assignment and were able to correctly answer at least 6 out of 8 questions, demonstrating the effectiveness of the course material in building skills necessary for a practical application of the theory.

\subsection{Analysis}
Overall, the survey results support the claim that the course material and its delivery can be highly effective in teaching XAI techniques. Analysis of individual ratings showed that this course is especially useful in promoting understanding of the diverse array of XAI techniques, and their conceptual distinctions, advantages, and disadvantages. The respondents' answers to open-ended questions suggested that interactive and practical aspects of the course were important in the successful process of transferring XAI theory into practical skills. The open questions and codebooks have been shown to be important parts of the course, especially in combination with live tutorials and online classes, that were open for brainstorming and discussions. The survey results also suggest that the course can be effective in teaching XAI techniques to individuals having no or minimal experience and knowledge about XAI.

Although only a small sample of respondents reported having no practical and/or technical ML experience, their high evaluations of the course’s ability to promote understanding of XAI, their self-reported success in meeting the learning objective, as well as ability to complete the course assignment might suggest that this course content and structure can also be effectively applied to teach individuals with little ML knowledge. In this case the course could potentially be adapted by, for example, splitting the first lecture ‘Preface to ML’ into two more elaborate lectures, that would also cover very basic technical aspects of ML. This could be an interesting approach to test in the future studies.

Student performance and forum questions suggested that, besides the technical aspects, such as consideration of Python libraries updates, the course could be improved by providing more support for the output analysis and evaluation aspect of the XAI. Most of the students found this part most challenging. This is also reflected in a slightly lower overall score of the ability to meet evaluation learning objectives, i.e., the ability to critically reflect on the results from the XAI techniques and suggest how it helps the problem context. It could potentially be due to the unfamiliar context/dataset, that is not relevant to their job or research interests but could also mean that more practical exercises followed by discussion of the analysis part should be included in the course. Potentially more practice analysing XAI outputs could lead to a better understanding/performance.

\subsection{Limitations}
The scores of the questionnaire are self-reported and reflect the subjective evaluation of their own understanding of the course material by the respondents. The high evaluations of the course effectiveness were reflected in the objective assignment performances by the students. However, the assignment responses and forum questions were only used as a supporting material and have not been analysed in a systematic way. More robust analysis will be possible once the second assignment and the final assignment results are available, allowing to evaluate a broader sample of student responses, measure improvement throughout the course, and see variations of the performance in applying different XAI techniques.

Although this course has been delivered to the data science experts working within the banking sector, in this paper only the students' responses were analysed. This limitation prevented evaluation of the generalisability of the course effectiveness across different settings and expertise levels. In the future, further surveys will be conducted to assess the effectiveness of the course for the more experienced data science and ML experts strictly working in the professional setting.

\section{Related Work}



As mentioned, primary inspiration in the structuring of this course includes recent surveys \citep{belle2020principles,ARRIETAxai, molnar2019}. Other works which have discussed XAI approaches broadly include \citep{ChakrabortiSZK17} wherein the authors analysis how to format explanations based on the internal model of the human receiving the explanation. With respect to specific XAI techniques, we include LIME as a notable approach \citep{RibeiroSG16}. In our recent deployment of the course, focus was given to SHAP as a local and global explainability method \citep{shap}. For local explanations we focused on counterfactuals and include various approaches in their derivation \citep{counterfactual,10.1007/978-3-030-58112-1_31}. Anchors was another local explainability method which provided clear logical rules as explanations \citep{AAAI1816982}. Other methods discussed in the course included Deletion Diagnostics \citep{doi:10.1080/00401706.1977.10489493} and  Influential Instances \citep{10.5555/3305381.3305576} where focus is on the impact specific instances have on model performance, The singular model-specific technique covered in the course was InTrees\citep{intrees} which can only be applied to tree-ensemble methods. Explainaility methods which focused on visualization included PDP \citep{friedman2001} and ICE \citep{goldstein}, both cases the methods provide global explanations.


We note other discussions of pedagogical perspectives such as \citep{SeufertXAI}, where focus is given to tools for learning analytics rather than course design structure. The authors discuss the potential of learning analytics to increase the efficiency and effectiveness of educational processes and the ability to identify and support students at potentially at risk of failure. Similarly, other pedagogical perspectives foucs on reporting a complete tutorial for a given topic, for example machine learning \citep{MeloMascarenhasPaiva2018}. Other works which address pedagogy in the context of structuring teaching methods include \citep{IJISM104} where the authors discuss a pedagogical approach with respect to E-learning. A more analytical pedagogical investigation is found in \citep{Sasson}, here the authors discuss effective learning spaces by collecting data from three schools and gathering teacher feedback. The focus is not given to a specific topic rather the broader approach in effective teaching. In \citep{Ellery}, the authors articulate pedagogical approach in order to address the problem of academic plagiarism.


Technical tutorials of XAI include \citep{SamekTUTxai}, where the lecturers propose methods for explaining deep neural networks. In \citep{CamburuTUTxai}, lecturers focus on the technical aspects of natural language explanations.  We note another XAI course designed for graduate students \citep{lakkaraju_lage_2019}, which does cover similar topics to our pedagogical approach on XAI, however greater emphasis is given to technical discussions of recent contributions in the field.

\section{Conclusion}

In this report we presented a pedagogical approach for structuring a course on explainability in machine learning. The aim of designing the course was to provide a formal introduction to the field of explainability. As mentioned, the field of XAI continues to advance at a rapid pace, but we argue that the fundamental ideas governing interpretability will remain relevant to the field. It is presumed that with the growing deployment of machine learning models, stakeholders will expect clarity on the relationship between training data and the decision boundary, identification of influential points in the training data, analysis of counterfactual arguments and justifications for model decisions, explanations in the form of extracted rules defined by a reasonable degree of readability. Therefore, even if the methodology themselves advance technically, the overall narrative structure for explainability will remain the same with the newer techniques. Although there are many approaches to explaining black-box models, there is an argument to be made that a structured pipeline needs to be designed to make these techniques more accessible and which can produce an explanatory summary elaborating on the usefulness and drawbacks in the deployment of a given model. We designed the course primarily for industry professionals, data scientists, and students with a programming and data science background. While experience with machine learning is a prerequisite, we do include a short primer on machine learning as a review. We believe that by following the content recommendations of this course, one can gain a fairly comprehensive understanding of how one goes from structured data, to model training, to an understanding of what the model decision boundary looks like and more, considering the various XAI techniques available. In the context of applying explainability, a broader point can be made that there is an interactive nature to having a model be approved by the stakeholder, where some of the XAI techniques can prove useful in ones argument for the model. Finally, our preliminary investigations into the effectiveness of this course has shown that the design choices in our pedagogical approach have been reasonable. As discussed earlier, courses such as these support inclusivity, empowerment and responsible design, by expressly making XAI accessible.

\begin{acks}

Vaishak Belle was supported by a Royal Society University Research Fellowship. Vaishak Belle was also supported by a grant from the UKRI Strategic Priorities Fund to the UKRI Research Node on Trustworthy Autonomous Systems Governance and Regulation (EP/V026607/1, 2020-2024). The authors acknowledge support from EPSRC grants EP/T517501/1 and EP/S035362/1. The authors acknowledge the financial support received by NatWest Group. This work was carried out in collaboration with University of Edinburgh's Bayes Centre and NatWest Group. We are especially grateful to Peter Gostev from the Data Strategy and Innovation team as well as a wide range of teams throughout Data and Analytics function at NatWest Group who provided insights on industry use cases, key issues faced by financial institutions as well as on the applicability of machine learning techniques in practice.

\end{acks}

\bibliographystyle{ACM-Reference-Format}
\bibliography{test}

\appendix
\section{Questionnaires}


\includepdf[pages=-]{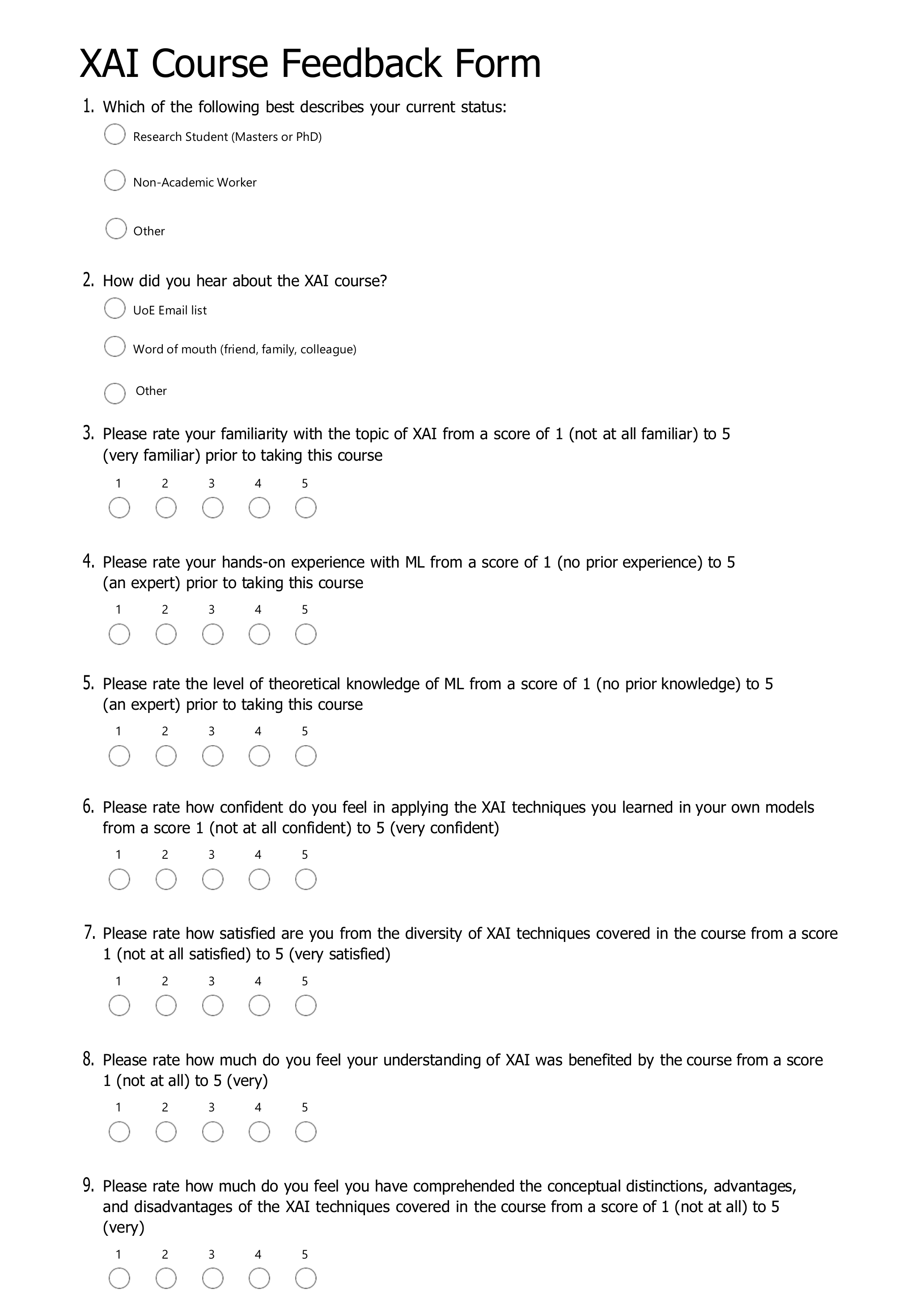}

\end{document}